\documentclass{article}

\usepackage{microtype}
\usepackage{graphicx}
\usepackage{amsfonts}
\usepackage{booktabs}
\usepackage{enumitem}
\usepackage{ulem}
\usepackage{listings}
\usepackage{xcolor}
\usepackage{subfig}
\usepackage{multirow}
\usepackage{amsmath}

\usepackage[accepted]{mlsys2026}
\usepackage{hyperref}

\newcommand{\OP}{ProTrain}

\newcommand{\new}[1]{{\color{black} #1}}

\newcommand{\newnew}[1]{{\color{black} #1}}

\mlsystitlerunning{ProTrain: Efficient LLM Training via Automatic Memory Management}

\begin{document}

\twocolumn[
\mlsystitle{ProTrain: Efficient LLM Training via Automatic Memory Management}

\mlsyssetsymbol{equal}{*}

\begin{mlsysauthorlist}
\mlsysauthor{Hanmei Yang}{to}
\mlsysauthor{Jin Zhou}{to}
\mlsysauthor{Yao Fu}{goo}
\mlsysauthor{Xiaoqun Wang}{goo}
\mlsysauthor{Ramine Roane}{goo}
\mlsysauthor{Hui Guan}{to}
\mlsysauthor{Tongping Liu}{to}

\end{mlsysauthorlist}

\mlsysaffiliation{to}{University of Massachusetts Amherst}
\mlsysaffiliation{goo}{Advanced Micro Devices, Inc.}

\mlsyscorrespondingauthor{Hanmei Yang}{hanmeiyang@umass.edu}

\mlsyskeywords{ML System, Memory Optimization, Data Parallelism, ZeRO, Gradient Checkpointing, Tensor Offloading}

\vskip 0.3in

\begin{abstract}
Memory pressure has emerged as a dominant constraint in scaling the training of large language models (LLMs), particularly in resource-constrained environments. While modern frameworks incorporate various memory-saving techniques, they often expose low-level configuration knobs that require manual tuning and specialized system expertise. This not only adds engineering overhead but also risks suboptimal hardware utilization when misconfigured.
This paper introduces \OP{}, a novel training system that automatically tailors memory management policies to the model architecture and underlying hardware resources, eliminating the need for manual intervention. The core of \OP{} is its automated memory management that abstracts complex memory management strategies into a few tunable configuration parameters, allowing searches for optimal parameter settings using cost models. \OP{} is equipped with a runtime profiler that provides precise estimates of latency, memory usage, and I/O bandwidth to build high-fidelity cost models. 
ProTrain does not change the training algorithm and thus does not compromise accuracy. Experiments show that ProTrain improves training throughput by 1.43$\times$ to 2.71$\times$ compared to the state-of-the-art training systems.

\end{abstract}
]

\printAffiliationsAndNotice{}

\section{Introduction}
\label{sec:intro}

Large Language Models (LLMs) have achieved remarkable success in various fields. 
Inspired by the scaling law~\cite{kaplan2020scaling} that LLM perplexity often improves logarithmically with the number of parameters, there has been a trend towards increasing parameter size. 
However, the significant growth in parameter size leads to a substantial increase in memory demands. According to existing studies~\cite{ren2021zero}, each unit increase in parameters generally requires $16\times$ more memory to store the model states (e.g., fp16 and fp32 parameters, fp16 gradients, fp32 momentum and variances), not to mention the increased memory demand for activations due to larger model sizes. Consequently, memory has become the dominant bottleneck in LLM training.

\textbf{Limitations of Existing Work}
Numerous memory management strategies have been proposed to alleviate the memory pressure, including parallelization~\cite{li2020pytorch, shoeybi2019megatron, huang2019gpipe}, gradient checkpointing~\cite{chen2016training, jain2020checkmate, herrmann2019optimal, zhao2023rockmate, korthikanti2023reducing}, and tensor swapping~\cite{rhu2016vdnn, le2018tflms, huang2020swapadvisor, ren2021zero, sun2022stronghold}.  
These approaches are often implemented in LLM training frameworks~\cite{rasley2020deepspeed, li2023colossal, zhao2023pytorch}, yet each exposes a set of low-level configuration knobs that require extensive manual tuning. 

The tuning challenge arises not only from the large number of parameters, but also from the lack of proper abstractions to manage their complex interdependencies.
First, \textit{techniques are often mutually exclusive in their applicability}. Consider activation memory management: checkpointing reduces memory usage by recomputing activations, while swapping trades I/O overhead for memory savings by offloading them to CPU. When computation can effectively hide this I/O overhead, swapping outperforms checkpointing, yet most practical systems favor checkpointing for its simplicity. The optimal choice thus depends on runtime dynamics rather than static heuristics.
Second, \textit{even complementary techniques compete for shared hardware resources}. For example, activation swapping and parameter prefetching both consume CPU-GPU bandwidth. When swapping is enabled to reduce activation memory, it can saturate the bandwidth, blocking parameter prefetches and stalling computation. Balancing these techniques requires coordinating their usage based on available hardware resources, yet such hidden dependencies make optimal configurations difficult to identify manually.
Together, these interactions create a search space that is both large and non-decomposable, making manual tuning infeasible without deep system expertise.

This challenge is particularly pronounced in systems such as DeepSpeed, which exposes over 18 tunable parameters spanning ZeRO partitioning, tensor swapping, and gradient checkpointing that are tightly coupled. For instance, \texttt{stage3\_max\_reuse\_distance} caches recently used parameters on the GPU to avoid reloading them during the backward pass, while \texttt{stage3\_max\_live\_parameters} limits the total number of parameters allowed to reside in GPU memory. This creates a conflict: aggressive caching consumes the available parameter budget, leaving insufficient room for prefetching upcoming parameters and delaying computation. Balancing these configurations requires runtime-aware tuning based on hardware resources, and the tight coupling among these knobs makes independent tuning ineffective, requiring joint exploration of an exponentially large search space.
Without such tuning, performance is suboptimal. Our evaluation shows that training a 10B GPT-2 model on RTX 3090 GPUs with default configurations utilizes only 35.6\% of GPU memory and runs $1.18\times$ slower than an optimized setup. Moreover, configurations optimized for RTX 3090 GPUs fail to fully utilize A100 GPUs, while those tailored for A100s often cause out-of-memory (OOM) errors on RTX 3090s, requiring users to manually re-tune for every hardware change.

\textbf{Our Contribution} To address these challenges, we present \OP{}, an efficient LLM training system that automatically tailors memory management policies \new{for data-parallel training} to the target model and hardware configuration. 
The core idea behind \OP{} is to unify disparate memory-saving techniques, such as ZeRO-style model state partitioning, tensor swapping, and gradient checkpointing, into a structured configuration space. Instead of exposing users to fragmented and low-level tuning knobs, \OP{} abstracts these strategies into a compact set of high-level parameters that capture the trade-offs and constraints across techniques. These parameters can then be automatically optimized using runtime and memory cost models informed by profiling.
\OP{} makes three core technical contributions:

\begin{itemize}
    \item \textit{Structured Memory Strategies:} We design two coordinated strategies that provide a unified view of memory management techniques. First, \textit{hierarchical chunk management} handles model states at both inter- and intra-chunk levels to promote memory reuse and overlap computation with communication. Second, \textit{interleaved block management} manages activations by interleaving tensor swapping and gradient checkpointing at the transformer block level, improving scheduling flexibility and minimizing memory fragmentation.
    \item \textit{Memory-Aware Profiler:} We introduce a memory-aware profiler that captures memory usage patterns often missed by conventional tools, such as transient allocations and memory from unhookable operations. The profiler reconstructs peak memory usage under varied scheduling policies and supplies detailed statistics on latency, memory footprint, and I/O bandwidth—providing critical inputs to our cost models.
    \item \textit{Automatic Memory Management:} We formulate memory management as a constrained optimization problem: minimize per-iteration runtime while staying within GPU memory capacity. Leveraging the structured search space and profiling data, \OP{} builds cost models that capture how different configurations affect runtime and memory usage. While memory optimization behaviors, such as asynchronous tensor swapping, are challenging to model due to runtime variability, we redesign the memory strategies to expose deterministic scheduling structures. This makes their interactions more predictable and enables the cost models to reason about compound effects, such as bandwidth contention and overlapping behaviors, ultimately guiding automatic and efficient configuration searches.
\end{itemize}

We evaluate \OP{} on popular LLM architectures including GPT-2, OPT, Mistral, and LLaMA, and compare against state-of-the-art frameworks such as DeepSpeed, Colossal-AI, and FSDP.
\OP{} trains models up to $2.47\times$ larger than DeepSpeed and $1.48\times$ larger than Colossal-AI on RTX 3090 GPUs, and up to $7.5\times$ larger than FSDP on A100 GPUs, while delivering $1.43 \times$ to $2.71 \times$ higher training throughput across model and hardware configurations.
In addition to throughput, \OP{} demonstrates strong scalability with available compute resources and consistently outperforms existing frameworks across a range of batch sizes. Its runtime and memory estimators achieve prediction errors under 4\%, enabling reliable automatic strategy selection.
These results highlight \OP{}'s effectiveness in delivering efficient and scalable memory management for LLM training under diverse deployment scenarios.

\section{Background and Related Work}
\label{sec:background}

Model training consists of three key stages: forward propagation (FWD), backward propagation (BWD), and parameter updates (OPTIM). LLM training typically employs mixed-precision training~\cite{micikevicius2017mixed}, where FWD and BWD are performed in reduced-precision formats (e.g., fp16 or bf16), while parameter updates use higher precision (e.g., fp32) to ensure accuracy.
Memory consumption during training arises from two main sources: \textit{model states} and \textit{residual states}. Model states include parameters, gradients, and optimizer states, while residual states consist of activations and temporary tensors. 

Among the many memory-saving techniques developed for LLM training, this work focuses on three widely adopted approaches.  
(a) \textit{Zero Redundancy Optimizer (ZeRO)}, a representative of sharded data parallelism, partitions model states across GPUs to eliminate redundancy and utilize aggregate device memory. We adopt this strategy as the parallelism foundation of our system due to its simplicity and widespread use.  
(b) \textit{Gradient checkpointing} reduces activation memory usage by discarding selected tensors during the forward pass and recomputing them during the backward pass.  
(c) \textit{Tensor swapping}\footnote{As we consider swapping to CPU memory, we use \textit{swapping} and \textit{CPU offloading} interchangeably.} offloads tensors, either model states or activations, to external memory such as CPU DRAM or NVMe devices.  
These techniques target different tensor types and offer complementary trade-offs between memory efficiency and computation overhead.
A more detailed discussion of related work is provided in Appendix~\ref{sec:relatedwork}.

\section{Overview of \OP{}}
\label{sec:design}

\begin{figure*}[t]
\begin{center}
\includegraphics[width=0.7\textwidth]{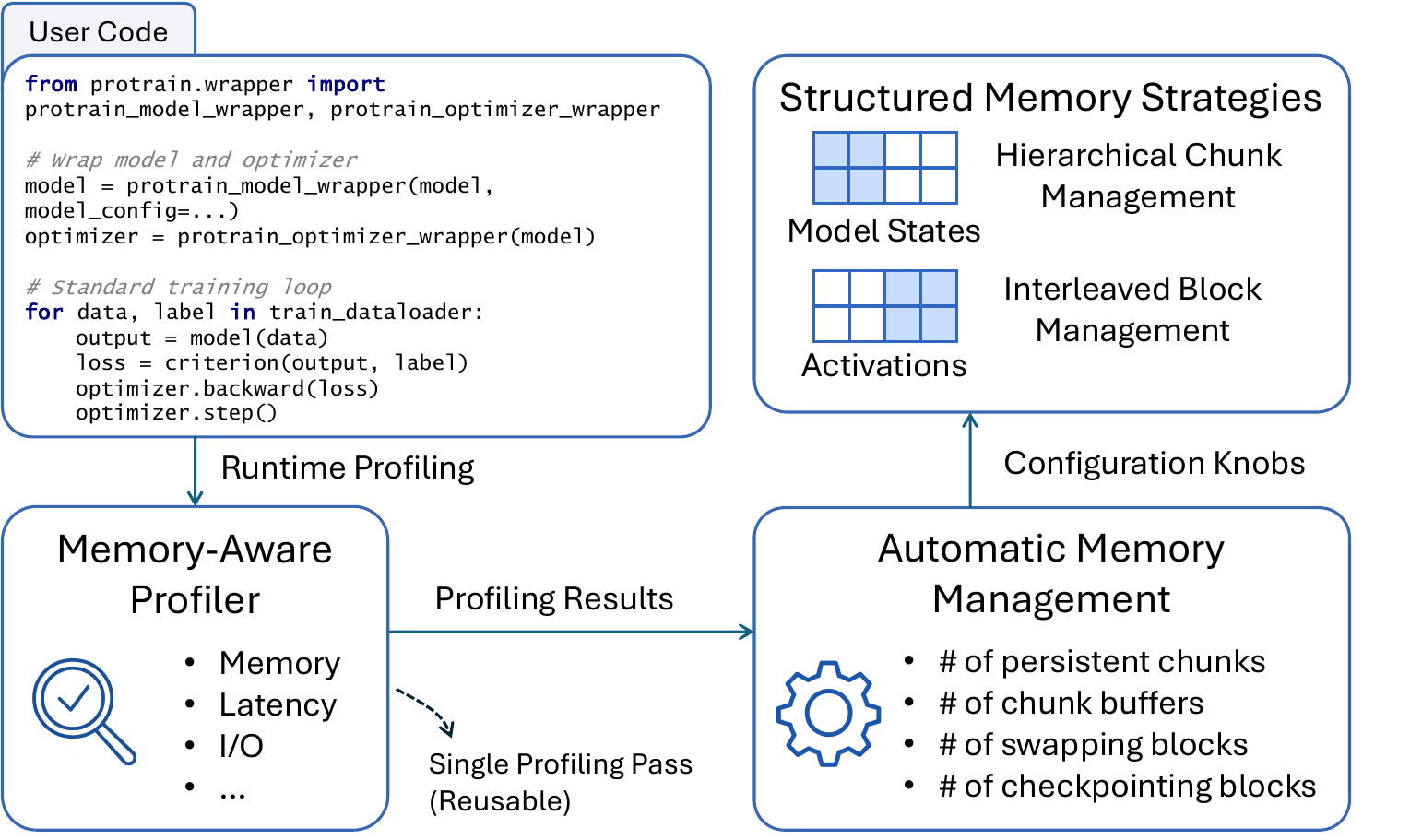}
\end{center}
\caption{System overview of \OP{}.}
\vspace{-0.1in}
\label{fig:protrain}
\end{figure*}

\OP{} comprises three core components, as illustrated in Figure~\ref{fig:protrain}: 
(1) \textbf{Structured Memory Strategies} (Section~\ref{sec:memory-policy}), which manage model states and activations using strategies suited to their specific optimization techniques.
(2) a \textbf{Memory-Aware Profiler} (Section~\ref{sec:profiling}) that collects runtime and memory data to guide memory management decisions, and
(3) an \textbf{Automatic Memory Management} module (Section~\ref{sec:management}) that dynamically identifies the optimal memory optimization strategies for training the target LLM on the given hardware.

\subsection{Structured Memory Strategies}
\label{sec:memory-policy}
 
\OP{} applies different strategies to manage model states and activations based on their unique optimization needs.
For model states, ZeRO and offloading are often used to alleviate GPU memory pressure. Activations, on the other hand, can be handled via gradient checkpointing and offloading to reduce memory footprint. 
These strategies unify the configurations of ZeRO, offloading, and checkpointing through a compact and structured design that enables predictable scheduling and effective automated tuning.
In the following, we elaborate the specific approaches for managing them.

\subsubsection{Hierarchical Chunk Management for Model States}
\label{sec:model-state-chunk}

\OP{} proposes a \textbf{hierarchical chunk management} system to manage model states that include parameters, gradients, and optimizer states. 
The system borrows the idea of organizing model states into chunks from prior work~\cite{fang2022parallel, li2023colossal, zhao2023pytorch}, enabling grouped transfers to improve I/O throughput.
However, existing chunk-based systems face the following limitations:

\begin{itemize}[noitemsep, nolistsep]
\item \textbf{Ineffective Overlapping:} The transfer of model states may not effectively overlap with computation if all model states have to be transferred. In addition, inefficient use of heterogeneous hardware leaves some overlapping opportunities unexploited.

\item \textbf{Ping-Pong Access Pattern:} Existing systems organize chunks by model-defined order rather than execution order, resulting in back-and-forth access that increases peak memory usage, causes cache inefficiencies, and triggers unnecessary transfers due to evictions.

\item \textbf{Frequent Memory Allocations: } Existing systems typically allocate memory dynamically to hold chunks. This approach fails to exploit the repetitive patterns of the training process, resulting in unnecessary allocation overhead and increased memory fragmentation.

\end{itemize}

\textit{\OP{}} overcomes these limitations through a two-level hierarchical approach: managing model states at both inter-chunk and intra-chunk levels. 
At the inter-chunk level, \textit{\OP{} addresses the ineffective overlapping issue by dividing model states into persistent and non-persistent chunks.}
\textit{Persistent chunks} reside on the GPU, eliminating data transfers and enabling direct GPU parameter updates.
\textit{Non-persistent chunks} are offloaded to CPU memory or other ranks, requiring parameters to be gathered on the GPU for computation and gradients to be reduced and offloaded back to the CPU for updates.
This design unifies ZeRO sharding and offloading: non-persistent chunks are partitioned to reduce communication overhead, while persistent chunks remain unpartitioned for fast GPU access. By adjusting the number of persistent chunks through its Automatic Memory Management module, \OP{} achieves a simple yet effective trade-off between memory usage and communication cost, without requiring complex coordination.

To further improve overlapping efficiency, \OP{} overlaps CPU parameter updates for non-persistent chunks with GPU backward computations, making effective use of idle CPU cycles.
Once the number of persistent chunks is determined, \textit{\OP{} allocates them sequentially from the beginning of the model}, while the remaining chunks are marked as non-persistent.
This strategy is motivated by two observations: (1) The forward pass computation can often hide the overhead of model states prefetching for later layers, but not the first few layers. (2) The first few layers, being processed last during the backward pass, do not have enough time to fully hide the latency of CPU parameter updates.

At the intra-chunk level, \textit{\OP{} addresses the ping-pong access issue by organizing model states within each chunk according to their execution order.} 
By leveraging the repeatable access patterns in training, \OP{} registers runtime hooks for forward and backward computations to capture operator execution order, eliminating back-and-forth transfers and releasing memory earlier to prefetch future chunks.

\textit{To reduce memory allocations, \OP{} introduces pre-allocated chunk buffers for non-persistent chunks,} allowing upcoming chunks to be prefetched ahead of computation. As parameters within each chunk are arranged in execution order, evictions triggered by buffer shortages follow a predefined sequence, eliminating the need for complex dynamic eviction policies. 
With pre-allocated buffers and deterministic prefetching, \OP{} exhibits stable runtime behavior, enabling accurate modeling of both execution overhead and peak memory usage (Section~\ref{sec:management}).
Moreover, non-persistent chunks that remain in the buffer after the forward pass can be reused during the backward pass if not evicted, eliminating the need for reloading and enhancing performance.

\subsubsection{Interleaved Block Management for Activations}

Activation memory usage can be greatly reduced using techniques such as tensor swapping and gradient checkpointing. While these approaches are widely employed in model training, their application to LLMs presents distinct challenges:

\begin{itemize}[nolistsep, noitemsep]
    \item \textbf{Granularity Selection: } Existing approaches often rely on either coarse-grained techniques, such as uniformly applying gradient checkpointing to all transformer blocks~\cite{li2023colossal, rasley2020deepspeed}, which can underutilize GPU memory and unnecessarily slow down training, or fine-grained tensor-level management~\cite{Peng2020Capuchin, beaumont2021efficient}, which adds implementation complexity and scalability challenges due to the large search space created by the numerous activation tensors in LLMs.
    \item \textbf{Efficiency in Scheduling: } Even with appropriate granularity, inefficient scheduling between activation swapping and gradient checkpointing can result in higher memory requirements and degraded performance. 
\end{itemize}

\begin{figure}[t]
\begin{center}
\includegraphics[width=0.5\textwidth]{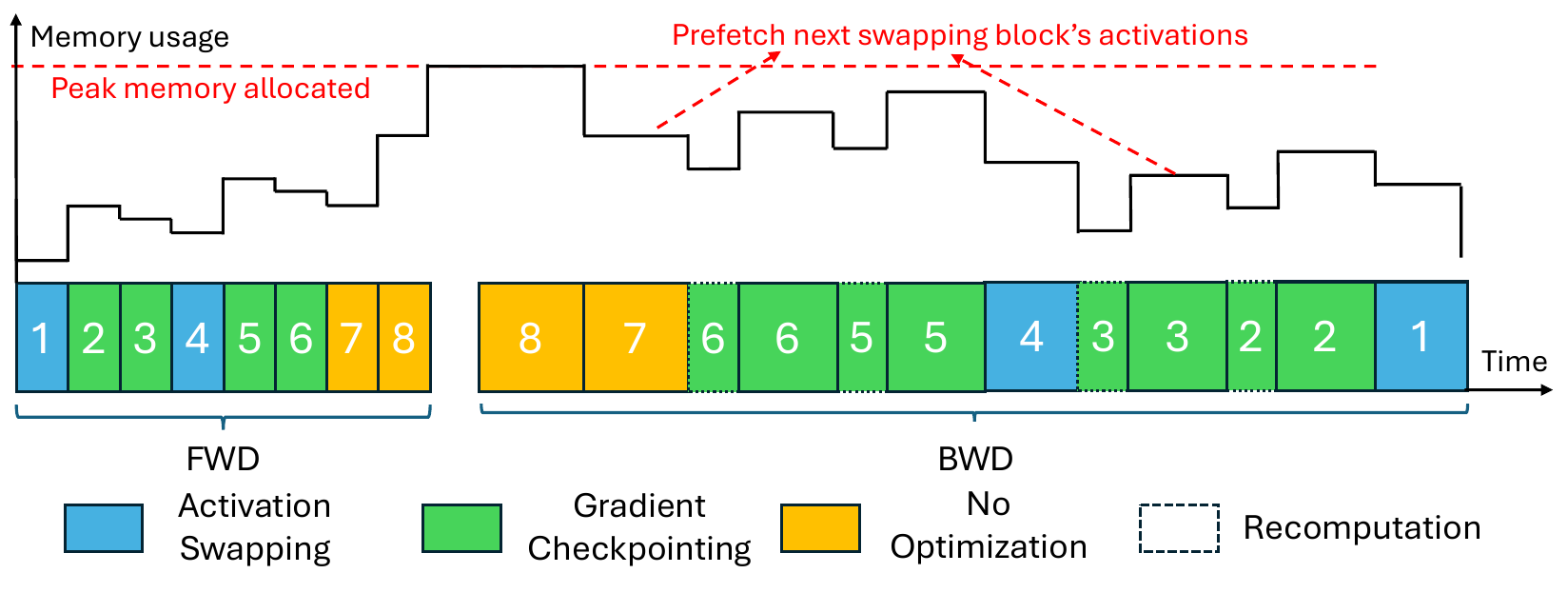}
\end{center}
\caption{Interleaved Block Management Layout and Memory Usage Trend}
\vspace{-0.2in}
\label{fig:block-mgr}
\end{figure}

To balance flexibility and scalability, \textit{\OP{} proposes a block-wise activation management strategy that applies memory optimization techniques at the transformer block level.}
Unlike coarse-grained strategies that enforce a uniform policy across all blocks, this approach allows each block to adopt the most suitable optimization scheme.
Each block can choose among three activation handling strategies: activation swapping, gradient checkpointing, or no optimization (i.e., neither swapping nor checkpointing). 
Compared to tensor-level management, the proposed block-wise approach adapts to existing transformer implementations without requiring computation graph reconstruction. It also significantly reduces the search space and scheduling overhead of managing individual activation tensors, offering a more scalable solution for LLMs.

\textit{\OP{} introduces an interleaved block management layout to coordinate activation swapping and gradient checkpointing, } as illustrated in Figure~\ref{fig:block-mgr}. The figure shows an 8-block transformer where blocks 1 and 4 use activation swapping, blocks 2, 3, 5, and 6 use gradient checkpointing, and the remaining blocks are left unoptimized.
This layout offers several benefits: (1) Placing swapping blocks earlier increases opportunities to overlap swapping with computation. (2) Interleaving swapping and checkpointing blocks prevents activation accumulation, reducing OOM risks caused by slower swapping and minimizing peak memory usage, as shown in the upper part of Figure~\ref{fig:block-mgr}. (3) Placing unoptimized blocks in later layers ensures their activations are consumed earlier, enabling timely prefetching for swapping blocks.

\subsection{Memory-Aware Profiler}
\label{sec:profiling}

\OP{} introduces a memory-aware profiler to determine the optimal settings for the tunable parameters defined in Section~\ref{sec:memory-policy}. The profiler focuses on estimating peak memory usage and understanding the runtime behavior of the model. In addition, it separately collects hardware metrics, such as memory transfer bandwidth and collective communication operation durations. The runtime profiler utilizes a single GPU to perform sampling under the user-defined batch size, which has the following challenges:

\begin{itemize}[noitemsep, nolistsep]
\item \textbf{Underestimation of Memory Usage: } Traditional profiling techniques, including static profiling~\cite{patil2022poet} and layer-wise runtime profiling~\cite{beaumont2021efficient}, fail to capture memory usage \new{from two sources: (1) transient memory for temporary tensors, which are intermediate results created within an operator that increase the peak memory footprint, and (2) memory usage from unhookable operators, typically PyTorch operators invoked via the functional API calls (e.g., \texttt{nn.functional.softmax} or \texttt{nn.functional.layer\_norm}) instead of \texttt{nn.Module} instances and are thus bypassed by layer-wise profiling hooks. These overlooked} components contribute about 17.2\% (3.06 GB) of peak memory in the 10B GPT-2 model with batch size 16. This underestimation leads to inaccurate memory planning and increases the risk of OOM errors during model execution.

\item \textbf{Accurate Profiling under Memory Constraints: } Profiling LLMs is inherently constrained by the limited memory capacity of a single GPU, often necessitating the use of memory-saving techniques to prevent OOM errors. However, these techniques change memory allocation patterns, making it difficult to accurately predict the peak memory usage of the model.

\end{itemize}

\textit{To accurately estimate peak memory usage, \OP{} proposes to profile operators within a complete model execution trace, rather than profiling them individually in isolation.}
This trace-based approach captures two types of memory deltas to handle both transient temporary tensors and unhookable operators.
The \textbf{intra-operator memory delta} measures the peak memory usage\footnote{ProTrain leverages PyTorch’s CUDA caching allocator statistics and uses pre/post operator hooks to track memory changes.} during an operator's execution minus the allocated memory just before it starts (i.e., the extra memory spike caused by intermediate results created within an operator). The \textbf{inter-operator memory delta} measures the peak memory usage between consecutive hooked modules minus the allocated memory at the end of the first hooked module, and is used to infer the memory cost of unhookable operators that fall between them. Together, these two measurements allow \OP{} to accurately estimate the full peak memory footprint of the model.

However, running a complete trace for large models is often infeasible on a single GPU. For example, a 70B model in FP16 requires 140 GB for model weights alone, exceeding the 80 GB capacity of an A100 GPU. To address this, \OP{} employs on-demand tensor management: tensors are allocated just before use and released immediately after, reducing peak memory to that of the largest single operator rather than the entire model.
\newnew{
This represents the most aggressive memory-saving strategy; in real training, however, model states and activations persist in GPU memory across operators, resulting in a much higher memory baseline. This raises the challenge of reconstructing the actual peak memory under any target configuration from a profiled trace that does not reflect real memory persistence patterns.
}

\textit{To enable peak memory estimation under altered usage patterns, we present a memory-aware profiling approach.} During profiling, \OP{} temporarily drops model states (i.e., parameters and gradients) and activations to reduce memory usage, as their fixed and predictable sizes allow their contributions to peak memory to be reconstructed through static analysis. 
Specifically, model states are pre-allocated in \OP{} and persist throughout training, contributing a constant memory cost that can be directly included in the estimation. In contrast, activations, managed by the training framework and tightly coupled to the operator execution order, require operator-by-operator analysis to quantify their impact on peak memory, as detailed in Section~\ref{sec:memory_cost_models}. By combining static analysis with operator-level memory tracking, our profiler provides accurate peak memory estimates across diverse configurations.

\subsection{Automatic Memory Management}
\label{sec:management}

\OP{} introduces an Automatic Memory Management module that dynamically optimizes memory configurations to adapt to different models and hardware setups, eliminating the need for tedious manual tuning while ensuring optimal efficiency. Automating memory management presents two key challenges:
\begin{itemize}[noitemsep, nolistsep]
    \item \textbf{Modeling runtime and memory usage}: Identifying optimal configurations requires precise modeling of how memory optimization strategies affect runtime and memory usage, which can be complex. For instance, while offloading reduces GPU memory demands, it introduces communication delays and intricate overlapping patterns that complicate runtime estimation. Predicting peak memory usage becomes difficult as swapping and checkpointing dynamically modify memory allocation patterns during execution.
    
    \item \textbf{Complications from Interactions Between Optimization Strategies}: Memory optimization strategies do not operate independently; rather, they influence and constrain each other. For example, parameter prefetching and activation swapping compete for bandwidth, potentially causing delays and inefficiencies when used together. Therefore, their interactions must be carefully evaluated rather than considered in isolation.
\end{itemize}

\textit{\OP{} addresses the modeling challenge by exploiting the structured search space defined by Structured Memory Strategies (Section~\ref{sec:memory-policy}).} The search space is abstracted into a compact set of tunable parameters, making their impact on runtime and memory usage more tractable to model.
Specifically, we abstract memory strategies into the following tunable parameters. For model states, we define (1) \textbf{chunk size} (${S_{\text{chunk}}}$) -- detailed in Section~\ref{sec:adaptivechunk}, (2) \textbf{number of chunks} ($N_{\text{chunk}}$), and (3) \textbf{number of persistent chunks} ($n_{\text{persist}}$) \textbf{and pre-allocated chunk buffers} ($n_{\text{buffer}}$) -- both bounded by $N_{\text{chunk}}$. 
For activations, we configure (1) \textbf{swapping interval} ($N_{\text{interval}}$) -- determined by the computation time required to swap out a block, (2) \textbf{number of blocks} ($N_{\text{block}}$), and (3) \textbf{number of blocks designated for activation swapping} ($n_{\text{swap}}$) \textbf{and gradient checkpointing} ($n_{\text{checkpoint}}$) -- both bounded by $N_{\text{block}}$.
The parameters ${S_{\text{chunk}}}$, $N_{\text{chunk}}$, $N_{\text{interval}}$, and $N_{\text{block}}$ are determined independently before the optimal configuration search. During configuration search, \OP{} tunes $n_{\text{persist}}$, $n_{\text{buffer}}$, $n_{\text{swap}}$, and  $n_{\text{checkpoint}}$ to identify the optimal setup.

\textit{\OP{} further formulates memory management as a constrained search problem that considers the interplay between different strategies.} The objective is to \new{minimize the per-iteration runtime}, while ensuring that peak memory usage remains within the available GPU memory capacity. To guide the search, \OP{} builds two cost models—one for runtime and one for peak memory usage—both defined as functions of the configuration parameters ${n_{\text{persist}}, n_{\text{buffer}}, n_{\text{swap}}, n_{\text{checkpoint}}}$. \new{Unlike approaches that require running actual training iterations for each candidate configuration, \OP{} builds these cost models from a single profiling pass and uses them to analytically evaluate all configurations in the search space, making the search tractable even when the configuration space is large. Detailed formulations are in Appendix~\ref{sec:runtime_cost_models} and Appendix~\ref{sec:memory_cost_models}. }

Building on the formulated cost models, \OP{} explores the configuration space through an exhaustive search while applying pruning strategies to improve efficiency. First, 
$n_{\text{swap}}$ is constrained by the swapping interval $N_{\text{interval}}$ and available bandwidth, which limits the number of feasible values to a small set. Second, configurations are evaluated in order of increasing memory usage, and those exceeding the GPU memory capacity are discarded early to avoid unnecessary evaluations. After traversing the search space, \OP{} selects the configuration parameters $\{n_{\text{persist}}, n_{\text{buffer}}, n_{\text{swap}}, n_{\text{checkpoint}}\}$ that achieves the shortest iteration runtime while satisfying memory constraints as the final setup.

\section{Implementation}
\label{sec:implement}

We implemented \OP{} on top of PyTorch with about 7,600 lines of code. It is designed for easy adoption and requires minimal changes to existing training code.
To use \OP{}, users only need to wrap the model and optimizer with the provided interfaces and the training loop remains unchanged. 
An example usage is shown in the top-left corner of Figure~\ref{fig:protrain}.
Internally, \OP{} uses runtime hooks to trace operator execution and tensor usage, enabling memory management without modifying the original model code. This design allows \OP{} to support any model that follows PyTorch’s standard forward API.
The detailed implementation components, including chunk organization and two low-level memory optimizations, are described in Appendix~\ref{sec:impl-details}.
\section{Evaluation}
\label{sec:evaluation}

\subsection{Experiment Setup}
\label{sec:setup}

\textbf{Workloads.}
We evaluate \OP{} on four popular LLM architectures: GPT-2~\cite{radford2019language}, OPT~\cite{zhang2022opt}, Mistral~\cite{jiang2023mistral}, and LLaMA~\cite{touvron2023llama}, using model implementations from the HuggingFace library. To cover a range of model sizes, we vary the hidden size, number of transformer blocks, and attention heads, as summarized in Table~\ref{tab:modelconfig}. The sequence length is fixed at 1024. \footnote{\new{We use 1024 for controlled comparisons, a standard setting when this study began. The methodology remains effective for longer sequences (e.g., the 75B model is trainable up to 7168 on an A100), as our search automatically adapts to the increased activation memory.}}

\begin{table}[ht]
\centering
\caption{Model configuration}
\resizebox{0.5\textwidth}{!}{
\begin{tabular}{lcccc}
\toprule
Model & Parameter Size & Hidden Size & \# of Blocks & \# of Heads  \\
\midrule
Mistral & 7B & 4096 & 32 & 32 \\
GPT-2 & 10B & 4096 & 48 & 32 \\
OPT, LLaMA & 13B & 5120 & 40 & 40 \\
GPT-2 & 15B, 20B, 30B, 40B & 8192 & 18, 24, 36, 50 & 64 \\
OPT & 30B & 7168 & 48 & 56 \\
LLaMA & 34B & 8192 & 48 & 64 \\
\bottomrule
\end{tabular}
}
\label{tab:modelconfig}
\end{table}

\textbf{Testbed.}
Experiments are conducted in two environments:
(1) \textbf{4$\times$ RTX 3090}: The system contains four NVIDIA GeForce RTX 3090 GPUs with 24GB memory. It is powered by Intel(R) Xeon(R) Silver 4214R CPU @ 2.40GHz with 24 cores. The CPU DRAM size is 384GB. The PCIe version is 3 with 15.8GB/s bandwidth. NVLink is not available in this setup.
(2) \textbf{4$\times$ A100}: The system contains four NVIDIA A100 GPUs with 80GB memory. It is powered by Intel(R) Xeon(R) Platinum 8480+ with 112 cores. The CPU DRAM size is 1TB. The PCIe version is 4 with 31.5GB/s bandwidth. GPUs are fully connected by NVLink 3.0 with 300GB/s bandwidth.

\textbf{Baselines.}
We compare \OP{} with three widely used open-source LLM training frameworks. \new{We focus on comparable features—ZeRO-3, CPU offloading, and gradient checkpointing. While implementations differ, these mechanisms provide similar functionality for fair comparison}: 
(1) \textbf{DeepSpeed}~\cite{rasley2020deepspeed} (v0.12.1): ZeRO-3 is enabled with offloading of both parameters and optimizer states. Thresholds such as \texttt{stage3\_max\_live\_parameter} and \texttt{stage3\_prefetch\_bucket\_size} are tuned for best performance. 
(2) \textbf{Colossal-AI}~\cite{li2023colossal} (0.3.3): We use the Gemini Plugin with chunk-based ZeRO-3 memory management. Both parameters and optimizer states are offloaded using a static placement strategy. 
(3) \textbf{FSDP}~\cite{zhao2023pytorch} (PyTorch 2.0.1): The native PyTorch implementation of ZeRO-3 with CPU offloading and selective gradient checkpointing. The \texttt{transformer\_auto\_wrap\_policy} is used to encapsulate each transformer block into a \texttt{FlatParameter}. 
Gradient checkpointing is either applied to all transformer blocks or entirely disabled, depending on whether memory constraints require it. For FSDP, we additionally evaluate selective checkpointing in Section~\ref{sec:selective}.

\subsection{Training Performance}

\begin{table}
    \centering
    \caption{Maximum trainable model size (unit: billion)}
    \resizebox{0.45\textwidth}{!}{%
        \begin{tabular}{lccccc}
        \toprule
        Backend & RTX 3090*1 & RTX 3090*4 & A100*1 & A100*4  \\
        \midrule
        \OP{} & \textbf{34B} & \textbf{37B} & \textbf{75B} & \textbf{87B} \\
        DeepSpeed & 15B & 15B & 34B & 37B \\
        Colossal-AI & 25B & 25B & 53B & 53B\\
        FSDP &  3B & 15B & 10B & 55B \\
        \bottomrule
        \end{tabular}
    }
    \label{tab:modelscale}
\end{table}

\subsubsection{Maximum Trainable Model Size}

Table~\ref{tab:modelscale} reports the maximum trainable model sizes across frameworks using GPT-2 as a benchmark.
\OP{} supports models up to 34B parameters on a single RTX 3090 GPU and scales to 37B with four GPUs. On A100 GPUs, \OP{} trains models as large as 75B on a single GPU and 87B with four GPUs, outperforming Colossal-AI and DeepSpeed by 1.64$\times$ and 2.35$\times$, respectively, in the four-GPU setup. In contrast, FSDP underperforms in the single GPU setting, handling significantly smaller models. Some frameworks also struggle to scale with multiple GPUs due to inefficiencies in model initialization.
These results demonstrate \OP{}'s ability to unlock larger model sizes on constrained hardware, making LLM training more accessible.

\begin{figure*}[!h]
  \centering
  \includegraphics[width=0.8\textwidth]{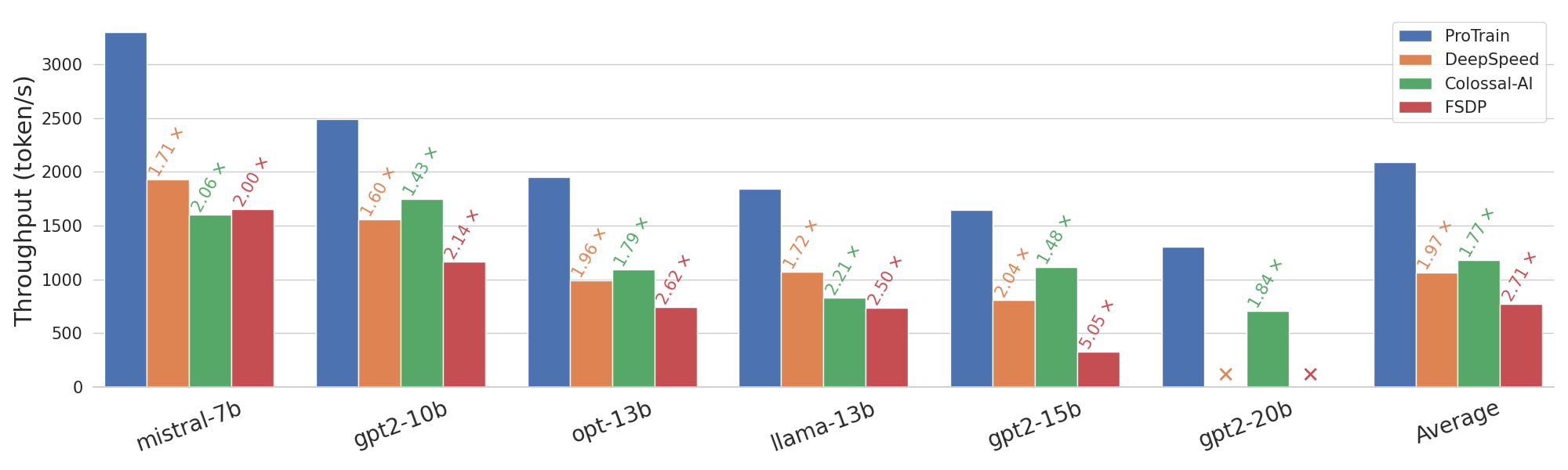}
  \includegraphics[width=0.8\textwidth]{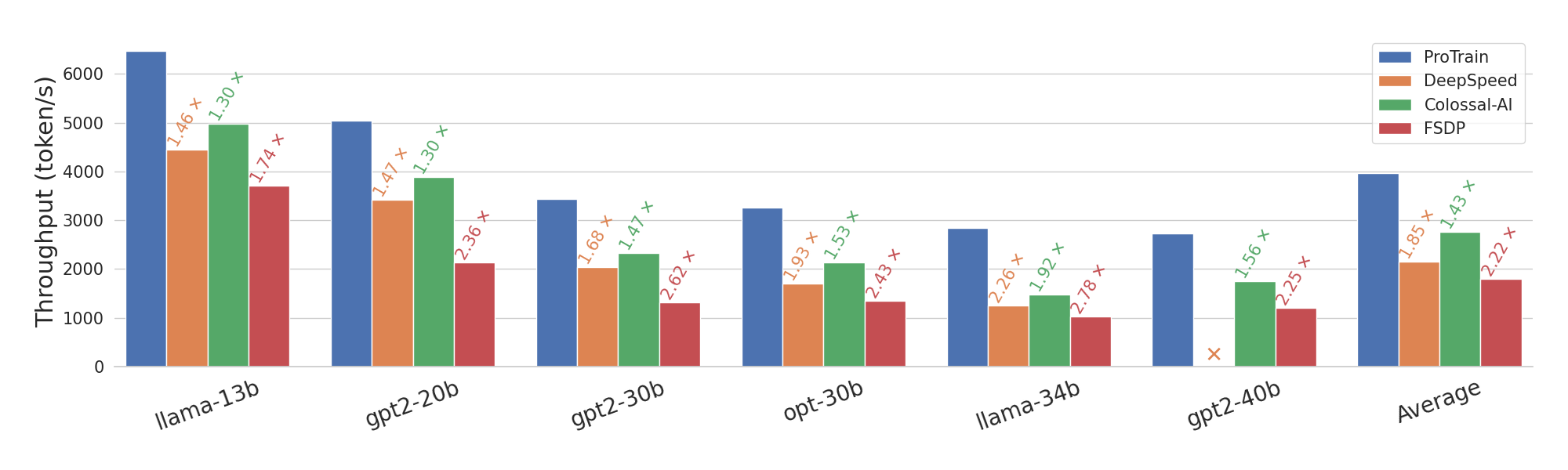}
  \caption{Maximum Training Throughput on four RTX 3090 GPUs (top) and A100 GPUs (bottom). The ``$\times$'' symbol indicates training failure due to Out-Of-Memory errors.}
  \label{fig:throughput}
\end{figure*}

\subsubsection{Training Throughput} Figure~\ref{fig:throughput} presents the maximum training throughput for various models on four RTX 3090 and A100 GPUs, measured in tokens per second. Throughput is obtained by testing each model at different batch sizes to find the highest achievable value. The results show that \OP{} consistently outperforms other frameworks across diverse hardware and models.
On RTX 3090 GPUs, \OP{} achieves an average throughput of 2090 tokens per second, 1.77 to 2.71$\times$ higher than other frameworks. On A100 GPUs, \OP{} improves the throughput of DeepSpeed, Colossal-AI, and FSDP by 1.85$\times$, 1.43$\times$, and 2.22$\times$, respectively.
While training throughput typically decreases as model sizes grow and memory demands increase, \OP{} maintains robust performance relative to other frameworks. For example, \OP{} achieves 5.05$\times$ the training speed of 15B GPT-2 on RTX 3090 GPUs and 2.78$\times$ that of 34B LLaMA on A100 GPUs compared to FSDP. In contrast, other frameworks often fail to train larger models with feasible batch sizes or resort to inefficient memory optimization techniques.
Overall, \OP{} delivers substantial performance improvements, achieving up to 2.71$\times$ the throughput of other frameworks on average and significantly enhancing the efficiency of LLM training.

\begin{figure*}[t]
    \centering
    \subfloat[]{{\includegraphics[width=0.4\textwidth]{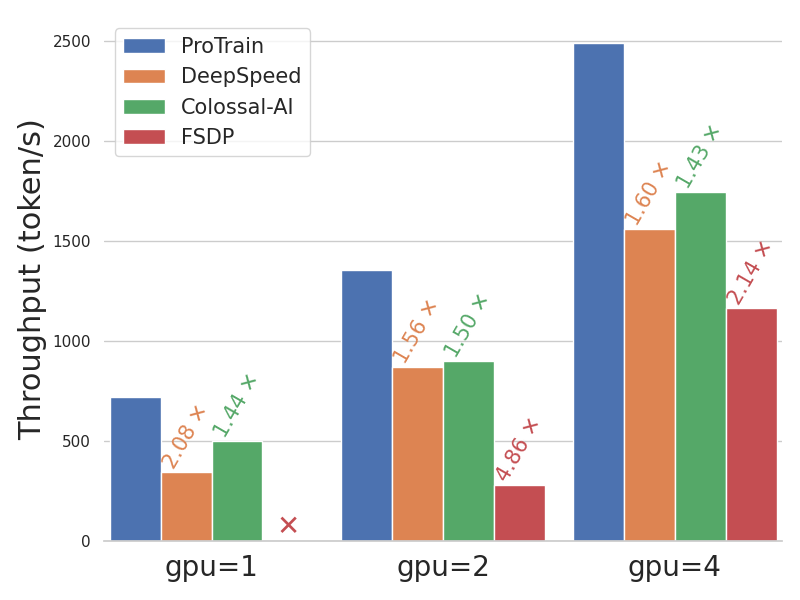} }}%
    \centering
    \subfloat[]
    {{\includegraphics[width=0.6\textwidth]{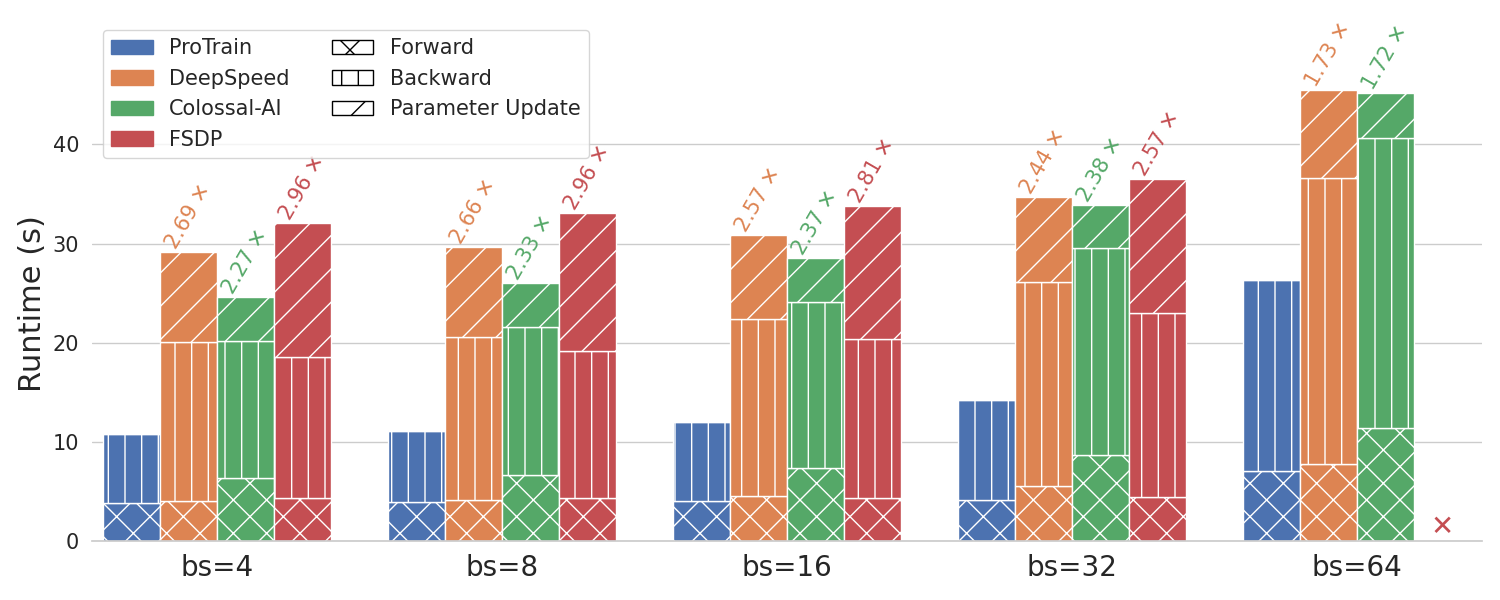} }}%
    \caption{Scalability of performance on RTX 3090 GPUs (a) Maximum throughput across different numbers of GPUs (b) Step time breakdown for different batch sizes}
    \label{fig:perfscale}
\end{figure*}

\subsubsection{Performance Scalability} 
Figure~\ref{fig:perfscale}(a) shows the maximum throughput of 10B GPT-2 with varying numbers of RTX 3090 GPUs. \OP{} scales efficiently, achieving 2493 tokens/s with four GPUs, a 3.5$\times$ improvement over the single-GPU baseline. While DeepSpeed and Colossal-AI also show increased throughput with more GPUs, their performance gains do not match those of \OP{}. 
Figure~\ref{fig:perfscale2}(a) further presents the scalability results for 34B LLaMA on four A100 GPUs. \OP{} continues to outperform the baselines, delivering a $2.49 \times$ to $3.58 \times$ speedup over a single GPU. This enhanced performance on A100s, compared to RTX 3090s, is largely due to \OP{}’s advanced memory management, which fully exploits the A100’s larger memory and higher bandwidth. As a result, \OP{} is able to scale to larger batch sizes and better utilize the available hardware resources to boost training throughput.

\subsubsection{Performance Breakdown}

Figure~\ref{fig:perfscale}(b) provides a detailed breakdown of one iteration time into forward, backward, and parameter update phases for training 10B GPT-2 across varying batch sizes on four RTX 3090 GPUs.
Detailed experimental results on A100 GPUs are provided in Appendix~\ref{sec:a100-scale}.

On RTX 3090 GPUs, \OP{} achieves notable speedups for two main reasons. First, it efficiently implements memory strategies through hierarchical chunk management and interleaved block management to optimize data movement and enhance I/O-compute overlap, with the benefits particularly evident during the backward pass due to its higher communication demands. Meanwhile, CPU parameter updates, executed concurrently with GPU backward computations, are effectively overlapped, making their contribution to the total iteration time nearly negligible. Second, \OP{}'s automatic memory management dynamically adjusts optimization strategies to balance performance and memory efficiency. As batch sizes grow, increasing memory demands prompt \OP{} to intensify memory-saving techniques, resulting in performance gains primarily driven by \OP{}'s efficient implementation.

\subsection{Ablation Study}
\label{sec:evalconfig}

\begin{figure}[!h]
    \centering
    \includegraphics[width=0.5\textwidth]{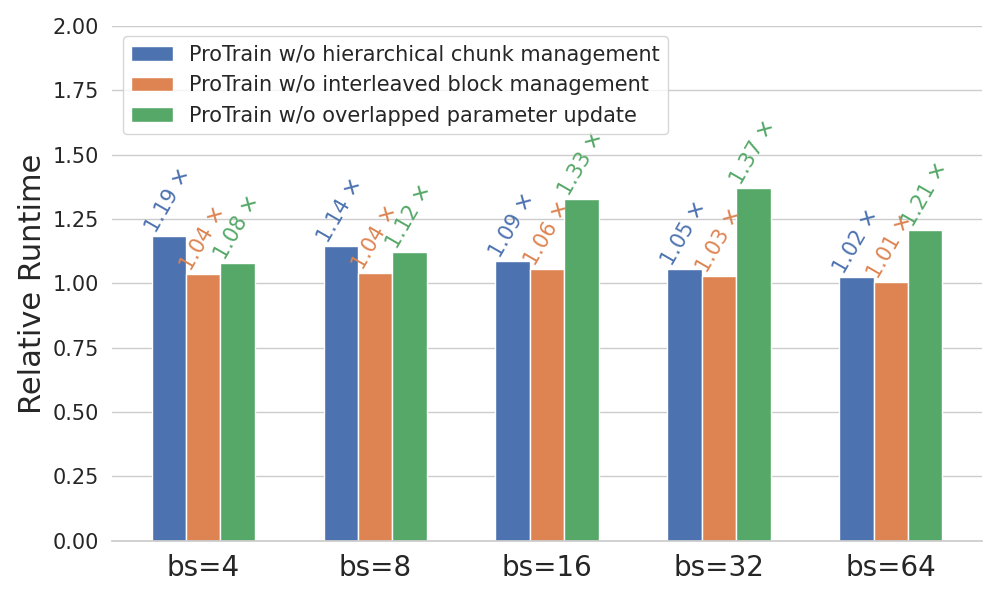}
    \caption{Impact of optimization strategies} 
    \label{fig:ablation}
\end{figure}

\subsubsection{Impact of Optimization Strategies} 
Figure~\ref{fig:ablation} analyzes the performance impact of disabling key optimizations in \OP{} when training a 10B GPT-2 model on four RTX 3090 GPUs. Each bar reports the runtime ratio of \OP{} without the corresponding optimization to that with the optimization enabled.
Disabling hierarchical chunk management, where persistent chunks are replaced by three chunk buffers, leads to a slowdown of $1.02\times$ to $1.19\times$. 
The importance of overlapped parameter updates grows with batch size, as larger batches require more parameters to be offloaded to the CPU. When overlapping is enabled, these CPU updates are effectively hidden behind GPU computations, improving overall efficiency; disabling this overlap, however, prevents such hiding and results in an average slowdown of $1.22 \times$.

Removing interleaved block management and instead applying gradient checkpointing to all transformer blocks results in an average slowdown of $1.04 \times$. 
\new{The limited gain in this setting is expected. On RTX 3090 GPUs with PCIe 3.0 bandwidth of 15.8 GB/s, training is communication-bound, and the available bandwidth is already saturated by parameter prefetching, leaving little room for activation swapping to overlap with computation. As a result, the automatic search favors checkpointing over swapping, producing a configuration already close to the ablation baseline of applying checkpointing to all blocks. 
On hardware with higher CPU-GPU bandwidth, such as the GH200 with NVLink-C2C (450 GB/s, $\sim$14.3$\times$ higher than A100 with PCIe 4.0), the relative cost of swapping decreases substantially. For example, for a 40B GPT-2 model, swapping activations per block takes 1.9s on A100 but an estimated 0.13s on GH200, while recomputation takes 0.95s on A100 and an estimated 0.3s on GH200, \newnew{scaled by their FP16 Tensor Core TFLOPS ratio}. This shift would lead \OP{} to favor swapping over checkpointing on GH200, yielding potentially larger gains from interleaved block management.

These results characterize the contribution of each optimization for a 10B GPT-2 model on RTX 3090 GPUs, and illustrate that their effectiveness varies across models and hardware. Section~\ref{sec:searched-config} further analyzes the optimal strategies under varying conditions, underscoring the necessity of \OP{}'s automatic approach.}

\begin{figure}[!h]
    \centering
    \includegraphics[width=0.5\textwidth]{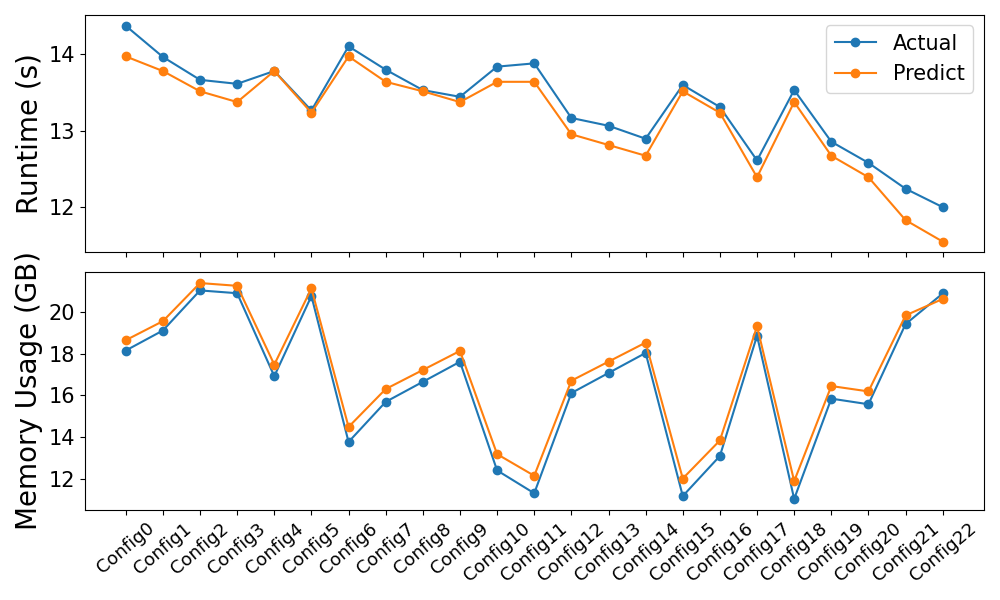}
    \caption{Predicted vs. actual runtime and peak memory across different configurations}
    \label{fig:autoconfig}
\end{figure}

\subsubsection{Effectiveness of the Runtime and Peak Memory Usage Estimators} 

Figure~\ref{fig:autoconfig} (top) compares the predicted and actual runtimes across a range of memory configurations when training a 10B GPT-2 model. The estimator not only closely tracks the overall runtime trend, but also achieves high accuracy, with deviations within 4\%. This fidelity enables \OP{} to reliably identify the configuration with the lowest runtime, demonstrating its robustness in navigating diverse memory strategies. 

Figure~\ref{fig:autoconfig} (bottom) shows the predicted versus actual peak memory usage across a wide range of configurations for training a 10B GPT-2 model. The estimator closely tracks the memory usage trend, with a maximum error within 4\%. 
Appendix~\ref{sec:eval-estimator} further validates the estimator’s reliability across different
models and batch sizes.

\subsubsection{Throughput Comparison with and without Offloading}
\label{sec:offload}

\begin{table*}[ht]
\centering
\caption{Training throughput on four A100 GPUs with and without offloading (tokens/s)}
\begin{tabular}{lccccc}
\toprule
\multicolumn{2}{c}{Model}     & Mistral-7B         & GPT2-10B & LLaMA-13B & GPT2-20B    \\
\midrule
\OP{}       & automatic & 11060.92 & 8266.40 & 6471.32 & 5043.75  \\
\midrule
\multirow{2}{*}{DeepSpeed}   & w/ & 7708.30 (1.43$\times$) & 6447.70 (1.28$\times$) & 4446.43 (1.46$\times$) & 3420.90 (1.47$\times$)  \\
                             & w/o & 9748.03 (1.13$\times$) & 7320.50 (1.13$\times$) & 5234.92 (1.24$\times$) & OOM  \\
\midrule
\multirow{2}{*}{Colossal-AI} & w/       & 7279.76 (1.52$\times$) & 6848.47 (1.21$\times$) & 4980.91 (1.30$\times$) & 3892.95 (1.30$\times$)  \\
                             & w/o      & 8447.30 (1.31$\times$) & 7855.46 (1.05$\times$) & 4404.30 (1.47$\times$) & 2084.74 (2.42$\times$)  \\
\midrule
\multirow{2}{*}{FSDP}        & w/       & 5315.81 (2.08$\times$) & 4666.03 (1.77$\times$)  & 3715.12 (1.74$\times$) & 2136.16 (2.36$\times$)  \\
                             & w/o      &  OOM         & OOM & OOM & OOM 
\\
\bottomrule
\end{tabular}
\label{tab:offload}
\end{table*}

\OP{} integrates CPU offloading as one of the key techniques to improve scalability, which may raise concerns about its practicality in real-world deployments due to limited bandwidth. To address this, we compare \OP{} against existing frameworks that do not rely on offloading. As shown in Table~\ref{tab:offload}, frameworks such as DeepSpeed and Colossal-AI exhibit competitive performance without offloading on smaller models (e.g., 7B Mistral, 10B GPT-2). However, their advantages diminish on larger models, where limited GPU memory restricts the achievable batch size. 
For instance, on 13B LLaMA, Colossal-AI can support a batch size of up to 220 with offloading, but only 32 without it, resulting in a 13\% slowdown when offloading is disabled. In comparison, \OP{} supports a batch size of 228 through more efficient memory strategy implementation and better overlapping mechanisms, achieving a $1.3\times$ higher throughput than the best-performing Colossal-AI configuration.
These results demonstrate that \OP{} consistently outperforms all baselines, both with and without offloading. More broadly, they highlight that offloading, when properly integrated, can serve as a scalable and cost-effective alternative to parallelism, enabling larger batch sizes and higher throughput under fixed hardware budgets, without requiring additional compute nodes.

\subsubsection{Profiling and Search Overhead}

\OP{} performs efficient profiling with overhead that scales linearly with a single training iteration. For instance, profiling 7B Mistral with batch size 4 on RTX 3090 GPUs takes 3.09 seconds, while 20B GPT-2 requires 5.38 seconds. The optimal configuration search is even faster, averaging just 0.06 seconds on the same hardware. These results demonstrate the low profiling and search overhead of \OP{}, enabling fast identification of optimal configurations with minimal cost.

\subsection{Analysis of Searched Configurations}
\label{sec:searched-config}

\begin{table*}[h]
    \centering
    \caption{Automatically searched configurations with the best performance}
    \renewcommand{\arraystretch}{1.2}
    \setlength{\tabcolsep}{5pt}
    \begin{tabular}{lccccccccc}
        \toprule
        ID & Model & Batch Size & Hardware & $N_{\text{block}}$ & $n_{\text{checkpoint}}$  & $n_{\text{swap}}$ & $N_{\text{chunk}}$ & $n_{\text{persist}}$ & $n_{\text{buffer}}$ \\
        \midrule
        A & GPT2-1B  & 8  & 4 RTX 3090 or A100 GPUs & 32 & 0  & 0  & 12 & 12 & 0 \\
        B & GPT2-1B  & 64 & 4 RTX 3090 GPUs & 32 & 24 & 2  & 12 & 2  & 3 \\
        C & GPT2-1B  & 64 & 4 A100     GPUs & 32 & 0  & 0  & 12 & 12 & 0 \\
        D & GPT2-10B & 8  & 4 RTX 3090 GPUs & 48 & 48 & 0  & 49 & 3  & 46 \\
        E & GPT2-10B & 8  & 4 A100     GPUs & 48 & 0  & 0  & 49 & 15 & 3 \\
        \bottomrule
    \end{tabular}
    \label{tab:searched_config}
\end{table*}

Table~\ref{tab:searched_config} summarizes the optimal memory management configurations automatically selected by \OP{} for different models, batch sizes, and hardware setups. The following discussion examines how these factors influence configuration choices.

\textbf{Batch Size Impact.}
Increasing the batch size from 8 (row A) to 64 (row B) on RTX 3090 GPUs leads to the following configuration adjustments: $n_{\text{swap}}$ increases from 0 to 2, $n_{\text{checkpoint}}$ increases from 0 to 24, $n_{\text{persist}}$ decreases from 12 to 2, and $n_{\text{buffer}}$ increases from 0 to 3. These changes align with runtime execution patterns, where a larger batch size amplifies computational intensity, allowing more data loading to be fully hidden within the computation. For 1B GPT-2 with batch size 64, \OP{} requires only three chunk buffers, each dedicated to prefetching, execution, and offloading. Since parameter loading overhead is fully hidden within the computation, the training remains compute-bound, eliminating the need for additional chunk buffers. The remaining bandwidth is then allocated to activation swapping, and \OP{} selects two swapping blocks whose overhead is entirely overlapped with computation without disrupting parameter loading. If bandwidth is insufficient for activation swapping in other setups, \OP{} prioritizes $n_{\text{checkpoint}}$ to minimize memory usage while maintaining performance.

\textbf{Hardware Impact.}
When training 1B GPT-2 with batch size 8 (row A), both A100 and RTX 3090 GPUs have sufficient memory, eliminating the need for offloading or activation checkpointing, resulting in identical configurations. 
However, with batch size 64 (rows B and C), the increased memory demand necessitates offloading and checkpointing on RTX 3090, while A100 can handle the workload without these optimizations due to larger memory capacity. 
For 10B GPT-2 with batch size 8 (rows D and E), both GPUs face memory constraints but adopt different strategies based on runtime characteristics. RTX 3090 GPUs, lacking NVLink and experiencing communication bottlenecks in NCCL operations, apply checkpointing to all blocks to free up space for larger $n_{\text{persist}}$ and $n_{\text{buffer}}$, mitigating parameter gathering overhead that cannot be fully hidden by computation. In contrast, A100 GPUs leverage NVLink’s high bandwidth, making training compute-bound, which results in retaining all activations and using three chunk buffers as the optimal configuration.

\textbf{Model Size Impact.}
The table shows that different model sizes require different configuration parameters. Larger models generally require more offloading (smaller $n_{\text{persist}}$ and $n_{\text{buffer}}$, and larger $n_{\text{swap}}$) and more gradient checkpointing (larger $n_{\text{checkpoint}}$). These adjustments optimize memory usage, allowing efficient training and fine-tuning of larger models within hardware constraints.

\subsection{Multi-Node Scalability}

To assess the scalability of \OP{} in large-scale training, we conduct a multi-node experiment on a 175B GPT model with batch size 256, deployed on 16 A100 GPUs across 4 nodes connected via a 100Gb InfiniBand network. The intra-node setup mirrors the configuration described in Section~\ref{sec:setup}. \OP{} achieves a throughput of 2218.70 tokens/s, outperforming DeepSpeed by $2.58\times$ and Colossal-AI by $2.09\times$, while FSDP fails due to OOM errors.
These results demonstrate \OP{}’s ability to generalize across nodes and scale effectively to larger clusters.

\section{Discussion}
\label{sec:discussion}

While offloading can alleviate memory pressure, its practicality in real-world LLM training is often questioned due to potential data transfer overhead. As shown in Section~\ref{sec:offload}, we compare \OP{} against frameworks both with and without offloading. Unlike prior systems that rely on fixed or manually tuned thresholds, \OP{} automatically evaluates the cost-benefit trade-off and applies offloading only when it improves performance. This adaptive decision-making allows \OP{} to maintain high throughput and consistently outperform other frameworks across diverse hardware settings.

\new{\OP{} focuses on data-parallel training with ZeRO, which is complementary to parallelism-focused systems~\cite{zheng2022alpa, miao2023galvatron, lin2024nnscaler} that optimize parallelism strategies but largely overlook fine-grained memory management. 
A closely related system is Mist~\cite{zhu2025mist}, which co-optimizes parallelism (DP, TP, PP) jointly with memory optimizations including checkpointing, ZeRO, and offloading. Mist automates a broader scope than \OP{} by jointly searching parallelism strategies and memory optimizations. 
It relieves users from choosing the parallelism configuration, a step that \OP{} still requires. 
However, this broader search space comes with trade-offs. \OP{} takes a narrower but deeper approach. 
By fixing the parallelism strategy, it can afford more precise memory modeling at the operator level, capturing intra-operator temporary tensors and unhookable operators that symbolic or coarser-grained models overlook. 
This also yields a more tractable search space, enabling fine-grained memory optimization that would be prohibitively expensive in a joint parallelism-memory search. 
In addition, \OP{}'s memory management layer operates on local tensors within each rank and is orthogonal to parallelism decisions, meaning it can in principle be composed with parallelism planners like Mist to combine broad automation with deep memory optimization.}

\new{Beyond performance gains and the scope discussed above, this work offers a new} perspective on scaling LLM training. While recent efforts have largely focused on \new{scaling through additional hardware or parallelism}, we highlight that better utilizing existing CPU memory, an abundant yet often underused resource, can substantially increase model capacity and batch size without additional investment in large-memory GPUs. Given the significant cost gap between commodity CPU memory and high-end GPUs, \OP{} provides a practical path to maximizing training potential within a fixed infrastructure budget.

\section{Conclusion}
\label{sec:conclusion}

This paper introduced \OP{}, a novel training system that simplifies the training process through automatic memory management. 
\OP{} designs tailored memory strategies for model states and activations and abstracts them into configuration parameters. It highlights the significance of precise memory usage and runtime data gathered through memory-aware profiling to build high-fidelity cost models that drive automated configuration search.
Through these innovations, \OP{} achieves up to 5$\times$ the performance of state-of-the-art frameworks and enables training models with up to 75 billion parameters on a single A100 GPU. 
We hope this work benefits AI researchers and practitioners by enabling more efficient use of existing hardware, supporting the scaling of LLM training without proportional increases in infrastructure cost.

\bibliography{ref}
\bibliographystyle{mlsys2026}

\appendix
\newpage
\clearpage

\section{Cost Models}

\subsection{Modeling Runtime}
\label{sec:runtime_cost_models}

We formulate automatic memory management as a constrained optimization problem and aim to minimize the total runtime of the training process. Since training consists of repeated iterations, minimizing the total training time is equivalent to minimizing the runtime of a single iteration, denoted as $T_\text{Iteration}$:

\begin{equation}
    \min_{\textit{configs}} T_\text{Iteration} \quad
    s.t. \; M_\text{Peak} < M_\text{Capacity},
\end{equation}
where $M_\text{Peak}$ represents the peak memory usage, and $M_\text{Capacity}$ is the total GPU memory capacity. The set of tunable configuration parameters, $\textit{configs}$, that determines the memory optimization strategy is $\{n_{\text{persist}}, n_{\text{buffer}}, n_{\text{swap}}, n_{\text{checkpoint}}\}$.

A single training iteration consists of three phases: the forward pass, the backward pass, and parameter updates. In \OP{}, forward and backward passes run on the GPU, while parameter updates can be performed on either the GPU or CPU. CPU updates execute concurrently with GPU computations, which include the backward pass and GPU-based updates. However, if CPU updates do not fully overlap with GPU operations, the total iteration time is constrained by the longer CPU update phase. The runtime cost model is formulated as follows:
\begin{equation}
    T_\text{Iteration} = T_\text{FWD} + \max\{T_\text{BWD} + T_\text{GPU\_OPTIM}, T_\text{CPU\_OPTIM}\}, 
    \label{eq:runtime}
\end{equation}
where $T_\text{FWD}$ and $T_\text{BWD}$ are modeled as a function of the configuration parameters. For parameter update of the persistent chunks ($T_\text{GPU\_OPTIM}$) and non-persistent chunks ($T_\text{CPU\_OPTIM}$), \OP{} models runtimes predictably based on parameter size.

To estimate the runtime of the forward pass, \OP{} adopts a chunk-based approach, as most operations operate at the chunk level. By comparing the computation and communication overheads for each chunk, the estimator identifies whether the chunk is compute-bound or communication-bound, using the larger value as its runtime estimate:

\begin{equation}
    T_\text{FWD} = \sum_{i=1}^{N_\text{chunk} + 1} \max \left( T_\text{comp}^\text{FWD}(i-1), T_\text{comm}^\text{FWD-prefetch}(i) \right),
    \label{eq:fwd}
\end{equation}

where $T_\text{comp}^\text{FWD}$ represents the forward computation time of a chunk, which aggregates the runtimes of individual operators within the chunk. $T_\text{comm}^\text{FWD-prefetch}$ represents the communication time required to prefetch parameters for the next chunk during the forward pass, which is calculated as follows:

\begin{equation}
T_\text{comm}^\text{FWD-prefetch}(i) =
\begin{cases}
0, & \text{if } i \leq n_{\text{persist}} \\
& \text{ or } i > N_{\text{chunk}}, \\
T_\text{comm}^\text{gather}(i) + T_\text{comm}^\text{upload}(i), & \text{otherwise},
\end{cases}
\end{equation}

where $T_\text{comm}^\text{gather}$ is the time to gather parameter chunks from multiple GPUs, and $T_\text{comm}^\text{upload}$ is the time to transfer non-persistent chunks from CPU to GPU. To estimate $T_\text{comm}^\text{gather}$ and $T_\text{comm}^\text{upload}$, \OP{} uses a separate hardware profiling to accurately model their runtime. In contrast to conventional approaches that assume a fixed bandwidth for memory transfers, \OP{} simulates various overlapping scenarios to capture the effects of bandwidth contention. For instance, when activation swapping is enabled, we estimate the swapping time, identify the affected chunks, and use the reduced bandwidth instead. The activation swapping time is excluded from the forward pass calculation, as \OP{} carefully controls $n_{\text{swap}}$ to ensure its overhead is fully overlapped with computation.

Similarly, the runtime of the backward pass is calculated at the chunk level:

\begin{equation}
\begin{aligned}
    T_\text{BWD} = \sum_{i=1}^{N_\text{chunk} + 1} \max ( T_\text{comp}^\text{BWD}(i) + T_\text{recomp}(i), \\
    T_\text{comm}^\text{BWD-prefetch}(i-1), 
    T_\text{comm}^\text{reduce-offload}(i+1) ).
    \label{eq:bwd}
\end{aligned}
\end{equation}

In contrast to the forward pass, the backward computation includes additional recomputation overheads from gradient checkpointing, represented by $T_\text{recomp}(i)$. 
The value is calculated as the aggregated forward computation time for the checkpointed blocks within chunk $i$, following the block-to-chunk mapping in the interleaved organization.
Another key distinction from the forward pass is the overhead related to gradient reduce and offloading during the backward pass, represented by $T_\text{comm}^\text{reduce-offload}$, which is defined as:

\begin{equation}
    T_\text{comm}^\text{reduce-offload}(i) =
    \begin{cases}
    T_\text{comm}^\text{reduce}(i), & \text{if } i \leq n_{\text{persist}}, \\
    0, & \text{if } i > N_{\text{chunk}}, \\
    T_\text{comm}^\text{reduce}(i) + T_\text{comm}^\text{offload}(i), & \text{otherwise}.
    \end{cases}    
\end{equation}

As with $T_\text{comm}^\text{FWD-prefetch}$, the performance of $T_\text{comm}^\text{reduce-offload}$ is directly influenced by the number of persistent chunks, as persistent chunks avoid parameter prefetching and only involve gradient reduce. However, $T_\text{comm}^\text{BWD-prefetch}$ differs in its estimation from $T_\text{comm}^\text{FWD-prefetch}$, and is defined as:

\begin{equation}
    T_\text{comm}^\text{BWD-prefetch}(i) =
    \begin{cases}
        0, & \text{if } i \leq n_{\text{persist}} \\
           & \text{ or } i > N_{\text{chunk}} - n_{\text{buffer}}, \\
        T_\text{comm}^\text{gather}(i) + T_\text{comm}^\text{upload}(i), 
           & \text{otherwise}.
    \end{cases}
\end{equation}

This difference arises because of the presence of chunk buffers, which cache the parameter loaded and gathered during the forward pass,  eliminating the need for re-loading and re-gathering in the backward pass. As a result, uploading and gathering are only required for chunks that were evicted due to limited buffer capacity.

Following the backward pass, parameter updates are executed on either the GPU or CPU, depending on the chunk type. \OP{} employs a fast CPU Adam optimizer~\cite{ren2021zero} for CPU updates and the FusedAdam optimizer~\cite{nvidia_apex_optimizers} for GPU updates. The runtime of both is modeled based on parameter size.

\subsection{Modeling Memory Consumption}
\label{sec:memory_cost_models}

Accurate peak memory estimation is essential for effective memory management, particularly in LLMs, where memory constraints require careful data handling to prevent exceeding capacity. 
Our estimator relies on the data collected by the profiler (detailed in Section~\ref{sec:profiling}) to compute memory usage precisely. The profiled data includes the changes in current memory usage,  $\Delta M_{\text{Cur}}^{\text{PriorOp}}$, and peak memory usage, $\Delta M_{\text{Peak}}^{\text{PriorOp}}$, before each operation, as well as $\Delta M_{\text{Cur}}^{\text{Op}}$ and $\Delta M_{\text{Peak}}^{\text{Op}}$ during each operation. Additionally, the profiler tracks the activation memory usage for each operator, $M_{\text{Act}}^{\text{Op}}$, and the memory usage at the end of the forward pass, $M_{\text{FWD}}$. Since memory usage typically peaks during the backward pass, our focus is on identifying the peak memory usage in that phase.

To estimate peak memory usage, we define two key variables: the current memory usage, $M_{\text{Cur}}$, and the peak memory usage, $M_{\text{Peak}}$. Initially, $M_{\text{Cur}}$ is set as:

\begin{equation}
\begin{aligned}
    M_{\text{Cur}} = M_{\text{FWD}} + \sum_{i=1}^{N_\text{op}} M_{\text{Act}}^{\text{Op}} (i)
    - M_\text{swap} \cdot n_\text{swap} \\
    - M_\text{checkpoint} \cdot n_\text{checkpoint}
\end{aligned}
\end{equation}

Here, $M_\text{swap}$ and $M_\text{checkpoint}$ represent the per-block memory savings from activation swapping and gradient checkpointing. Then, for each operator, we iteratively update $M_{\text{Cur}}$ and $M_{\text{Peak}}$ as follows:

\begin{equation}
\begin{aligned}
    M_{\text{Cur}}(i) &= M_{\text{Cur}}(i - 1) + \Delta M_{\text{Cur}}^{\text{PriorOp}}(i) + \Delta M_{\text{Cur}}^{\text{Op}}(i) \\
    &\quad - \begin{cases} 
        0, & \text{if } \text{Op}(i) \text{ uses swapping or checkpointing} \\ 
        M_{\text{Act}}^{\text{Op}}(i), & \text{if no optimization is applied}
    \end{cases}
\end{aligned}
\label{eq:cur-mem}
\end{equation}

\begin{equation}
\begin{aligned}
    M_\text{Peak} (i) = \max \big\{ & M_\text{Peak} (i-1), 
    M_\text{Cur} (i-1) + \Delta M_{\text{Peak}}^{\text{PriorOp}} (i), \\
    & M_\text{Cur} (i-1) + \Delta M_{\text{Cur}}^{\text{PriorOp}} (i) + \Delta M_{\text{Peak}}^{\text{Op}} (i) \\
    & + \mathbb{I}_{\text{checkpoint}}^{\text{Op}}(i) \cdot M_\text{checkpoint} \big\}.
\end{aligned}
\label{eq:peak-mem}
\end{equation}

where $\mathbb{I}_{\text{checkpoint}}^{\text{Op}}(i)$ is an indicator function set to 1 if the operator belongs to a checkpointing block and is the first operator in that block; otherwise, it is 0. This reflects the runtime behavior where gradient checkpointing triggers a recomputation before the actual backward computation, increasing peak memory usage. 
For swapping blocks, activation prefetching is only performed when sufficient memory is available, ensuring it does not contribute to peak memory, as illustrated in the memory usage trend of Figure~\ref{fig:block-mgr}.

This iterative, operator-wise approach estimates peak memory by accounting for transient temporary tensors and short-lived activations, closely mirroring their allocation and deallocation behavior during runtime. Finally, the overall peak memory considering model states is computed as:

\begin{equation}
    M_\text{Peak} = (M_\text{Peak} + M_\text{persist} \cdot n_\text{persist} + M_\text{buffer} \cdot n_\text{buffer}) \cdot \alpha
\end{equation}

where $M_\text{persist}$ and $M_\text{buffer}$ represent the memory allocated for a single persistent chunk and chunk buffer, respectively, while $\alpha$ is a factor that accounts for potential memory inefficiencies due to memory fragmentation.

\section{Implementation Details}
\label{sec:impl-details}

In the remainder of this section, we describe key implementation components, including chunk organization and size search (Section \ref{sec:adaptivechunk}) and two low-level memory optimizations (Section \ref{sec:memory-opt}) that improve memory efficiency. 

\subsection{Chunk Organization and Size Search}
\label{sec:adaptivechunk}

As discussed in Section~\ref{sec:model-state-chunk}, \OP{} organizes parameters according to their execution order rather than the initialization order used in Colossal-AI. 
For transformers that share parameters across layers, \OP{} uses the parameter's first occurrence as the ordering criterion. Additionally, \OP{} groups parameters from the same transformer block into one chunk, which minimizes memory accesses, especially when using gradient checkpointing that requires revisiting parameters in reverse during the backward pass. To identify the most efficient chunk size for a given model, \OP{} conducts a grid search that simulates memory waste across various chunk sizes and selects the one that minimizes it.

\subsection{Memory Optimizations}
\label{sec:memory-opt}

\paragraph{Single-Stream Memory Allocation}

\OP{} unifies memory allocations within the default stream to improve memory utilization. PyTorch's allocator adopts a multi-heap design where each stream has its own heap, limiting cross-heap memory reuse and necessitating the use of \texttt{record\_stream()} to ensure correctness. By using a single stream for all allocations and directly managing deallocation synchronization ourselves, we effectively prevent misuse and reallocation conflicts, thereby improving memory efficiency.

\paragraph{Customized Pinned Memory Allocator}

We observe that the default pinned memory allocator (\texttt{CUDAHostAllocator}) often over-allocates by rounding up to the nearest power of two, leading to significant memory waste. To address this inefficiency, \OP{} developed a customized pinned memory allocator that leverages insights from automatic memory management to precisely determine pinned memory requirements, providing finer control and avoiding the excessive memory reservation of the default allocator.

\section{Full Experiment Results}

\subsection{Throughput Scalability on A100 GPUs}
\label{sec:a100-scale}

\begin{figure*}[!h]
    \centering
    \subfloat[]{{\includegraphics[width=0.35\textwidth]{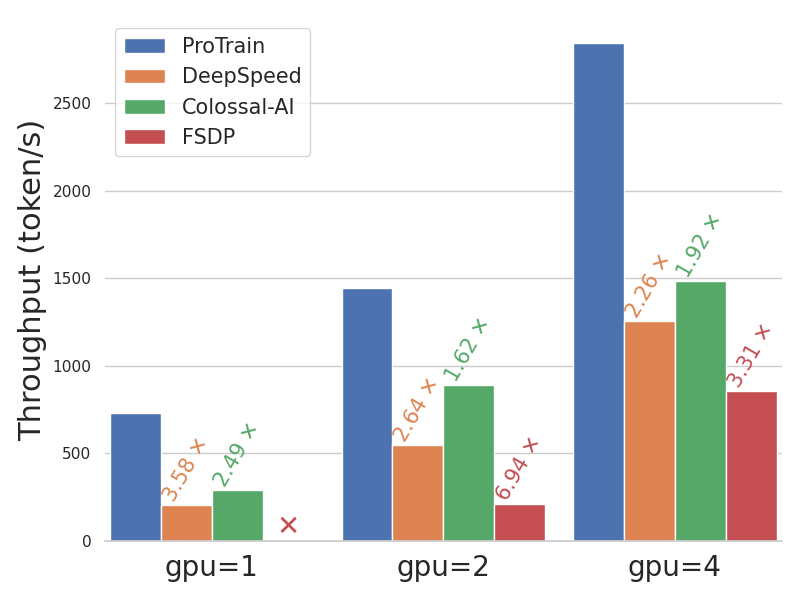} }}%
    \centering
    \subfloat[]
    {{\includegraphics[width=0.64\textwidth]{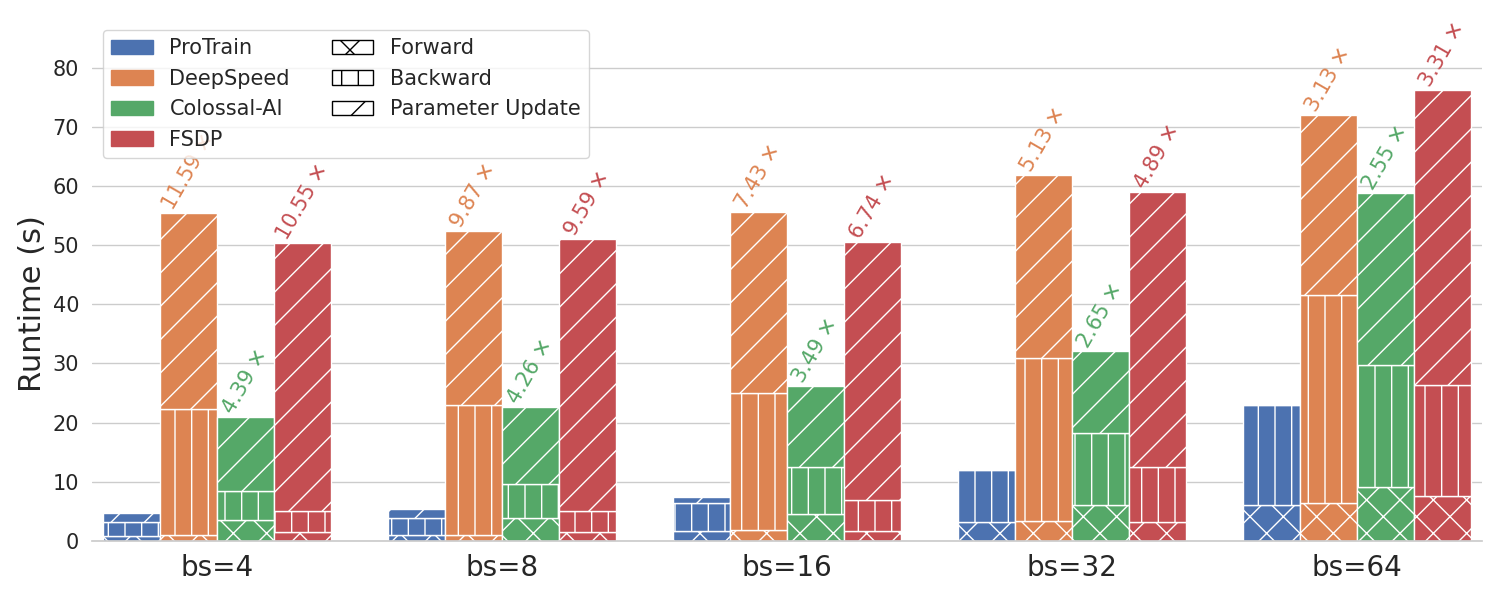} }}%
    \caption{Scalability of performance on A100 GPUs (a) Maximum throughput across different numbers of GPUs (b) Step time breakdown for different batch sizes}
    \label{fig:perfscale2}
\end{figure*}

Figure~\ref{fig:perfscale2}(b) breaks down the per-iteration runtime into forward, backward, and parameter update phases when training 34B LLaMA on four A100 GPUs.
On A100 GPUs, the performance advantage of \OP{} becomes more pronounced. It effectively overlaps CPU parameter updates with GPU computations, avoiding the update bottlenecks observed in frameworks such as FSDP, where the use of the default Adam optimizer leads to significant time spent in the update phase. 
Compared to DeepSpeed, \OP{} achieves lower backward pass latency by using fixed-size chunk buffers that enable timely parameter prefetching and reuse. In contrast, DeepSpeed relies on a threshold-based sliding window mechanism, where parameters are evicted only after their usage completes and new ones are prefetched only when enough memory is freed. This approach leads to poor bandwidth utilization and suboptimal overlapping.

\subsection{Effect of Runtime/Peak Memory Usage Estimator}
\label{sec:eval-estimator}

\begin{figure}[!h]
    \centering
    \includegraphics[width=0.5\textwidth]{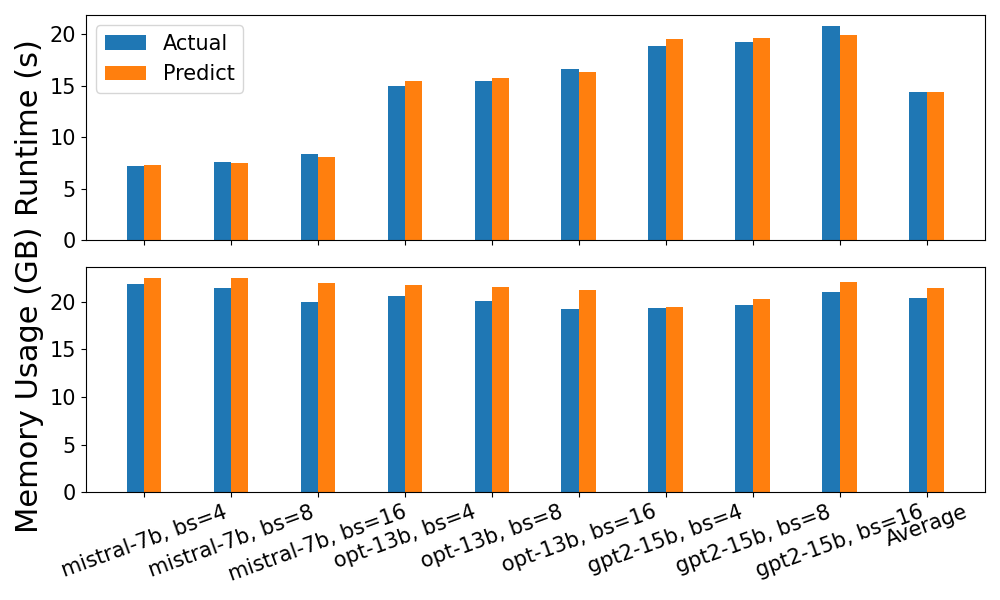}
    \caption{Predicted vs. actual runtime and peak memory usage across different models and batch sizes}
    \label{fig:estimator}
\end{figure}

To further validate its effectiveness, Figure~\ref{fig:estimator} (top) compares the predicted and actual runtimes of the configurations selected by \OP{} across different models and batch sizes on four RTX 3090 GPUs. Even across diverse workloads, the prediction error remains within 5\%, confirming the estimator's generalization capability. With precise runtime estimation, \OP{} can effectively automate the selection of memory management strategies tailored to each training scenario.

Figure~\ref{fig:estimator} (bottom) extends this evaluation to the best configurations selected by \OP{} across various models and batch sizes. 
Although larger batch sizes generally require more memory, \OP{} reduces persistent chunks and chunk buffers under tight budgets, resulting in lower overall memory usage.
The estimator conservatively overestimates peak memory by at most 10\%, effectively accounting for fragmentation and reducing the risk of OOM, thereby ensuring robust performance across varied training scenarios.

\subsubsection{Evaluation of FSDP with Selective Checkpointing}
\label{sec:selective}

\begin{table*}[h]
    \centering
    \caption{Training throughput of FSDP with and without selective checkpointing vs. ProTrain (tokens/s)}
    \begin{tabular}{lccc}
        \toprule
        Model & FSDP + Selective Checkpointing & FSDP - Selective Checkpointing & ProTrain \\
        \midrule
        LLaMA-13B & 3996.67 (1.00$\times$) & 3715.12 (0.93$\times$) & 6471.32 (1.62$\times$) \\
        GPT2-20B   & 2392.17 (1.00$\times$) & 2136.16 (0.89$\times$) & 5043.75 (2.11$\times$) \\
        GPT2-30B   & 1383.52 (1.00$\times$) & 1307.88 (0.95$\times$) & 3431.38 (2.48$\times$) \\
        OPT-30B   & 1621.85 (1.00$\times$) & 1342.40 (0.83$\times$) & 3266.02 (2.01$\times$) \\
        LLaMA-34B & 1247.25 (1.00$\times$) & 1024.23 (0.82$\times$) & 2845.18 (2.28$\times$) \\
        GPT2-40B   & 1143.06 (1.00$\times$) & 1208.68 (1.06$\times$) & 2723.50 (2.38$\times$) \\
        \bottomrule
    \end{tabular}
    \label{tab:fsdp_comparison}
\end{table*}

We re-evaluated FSDP using selective gradient checkpointing to compare its impact across hardware. On RTX 3090 GPUs, it does not improve throughput because execution is communication-bound, which limits the benefit of reduced recomputation. In contrast, on A100 GPUs, selective checkpointing improves throughput for all models except 40B GPT-2, which fails to scale due to OOM issues.
Table~\ref{tab:fsdp_comparison} reports the maximum throughput on A100 GPUs for FSDP with and without selective checkpointing, as well as ProTrain. While selective checkpointing improves FSDP’s performance, ProTrain consistently delivers higher throughput by jointly optimizing memory usage and hardware efficiency.

\section{Related Work}
\label{sec:relatedwork}

\paragraph{Tensor Swapping and Gradient Checkpointing}

Tensor swapping~\cite{rhu2016vdnn, le2018tflms, huang2020swapadvisor, ren2021zero, sun2022stronghold} expands memory capacity by offloading tensors to external memory such as CPU memory. Early work focused on activation swapping~\cite{rhu2016vdnn, le2018tflms}, while later efforts extended swapping to parameters~\cite{huang2020swapadvisor} and optimizer states~\cite{ren2021zero}.
Gradient checkpointing~\cite{chen2016training, jain2020checkmate, herrmann2019optimal, zhao2023rockmate, korthikanti2023reducing} reduces activation memory by recomputing intermediate results during the backward pass. 
Several studies~\cite{Peng2020Capuchin, beaumont2021efficient, nie2022tsplit} explore joint optimization of swapping and checkpointing, but primarily target convolutional networks and offer limited scalability to larger transformer models. In contrast, \OP{} integrates these techniques through transformer-aware abstractions and cost-driven coordination, enabling automatic and scalable memory management for large-scale model training.

\paragraph{Parallelization Techniques}

Parallelization is a widely adopted strategy to scale large model training by distributing computation and memory across multiple devices. Common approaches include: Data Parallelism (DP)\cite{li2020pytorch}, which replicates the model and distributes input data across devices, can be enhanced by ZeRO~\cite{rajbhandari2020zero, zhao2023pytorch} that partitions model states across GPUs to reduce memory redundancy; Tensor Parallelism (TP)\cite{shoeybi2019megatron, xu2021gspmd}, which partitions tensors within operators; and Pipeline Parallelism (PP)\cite{huang2019gpipe, narayanan2019pipedream}, which splits the model into stages assigned to different devices for pipelined execution. 
Recent systems~\cite{zheng2022alpa, lin2024nnscaler, miao2023galvatron} explore hybrid combinations of these parallelism strategies via automatic search.
\new{Mist~\cite{zhu2025mist} goes further by co-optimizing parallelism with memory footprint reduction techniques including checkpointing, ZeRO, and offloading; we provide a detailed comparison in Section~\ref{sec:discussion}.}
While effective, these systems focus purely on parallel execution and overlook memory-saving strategies such as swapping and checkpointing. Our work complements them by integrating these techniques into the search space, enabling further scalability in model size or throughput under fixed resources.
We focus on Data Parallelism with the ZeRO technique to demonstrate this integration in a practical and widely used setting, but our approach is orthogonal to other parallelisms and applicable to more complex training setups.

\paragraph{LLM Memory Optimization Techniques}

Beyond the aforementioned techniques, a range of system-level and algorithmic methods have been developed to further reduce memory consumption in LLM training. Compiler-level optimizations, such as operator fusion~\cite{zhao2022apollo, ansel2024pytorch}, eliminate intermediate tensors by merging operations during compilation. At the kernel level, memory-efficient attention mechanisms~\cite{dao2022flashattention, rabe2021self} reduce memory footprint by restructuring attention computation to avoid materializing the full attention matrix.
Quantization methods reduce memory usage by compressing weights, activations, and gradients into lower-precision formats such as INT8~\cite{dettmers2022gpt3}, FP8~\cite{fishman2024scaling, micikevicius2022fp8}, and FP4~\cite{yang2025microscaling}. 
While orthogonal to our approach, these techniques are complementary and can be integrated into \OP{} to further improve scalability.

\paragraph{Training Frameworks for Transformers}

To meet the growing demand for efficient LLM training, numerous frameworks have been developed with distinct design goals and optimizations. DeepSpeed~\cite{rasley2020deepspeed} by Microsoft enhances training efficiency through ZeRO series techniques~\cite{rajbhandari2020zero, ren2021zero, rajbhandari2021zero}, and supports a wide range of parallelism strategies, tensor swapping, and so on. Colossal-AI~\cite{li2023colossal} from HPC-AI Tech, offers similar features but distinguishes itself with a chunk-based memory management design~\cite{fang2022parallel}, which is also adopted in \OP{}. Megatron-LM~\cite{shoeybi2019megatron} by NVIDIA, on the other hand, focuses on model parallelism techniques. 
In parallel, recent academic efforts~\cite{sun2022stronghold, li2022harmony, feng2023mobius} aim to enable efficient training on memory-constrained systems. \OP{} complements these frameworks by introducing automatic memory management that adapts to varying model and hardware configurations.

\end{document}